\documentclass[runningheads]{llncs}

\usepackage{graphicx}
\usepackage{amsmath,amssymb} 
\usepackage{color}
\usepackage{graphicx}
\usepackage{tabulary}
\usepackage{enumitem}
\usepackage{multirow}
\usepackage{soul}

\newcolumntype{K}[1]{>{\centering\arraybackslash}p{#1}}

\DeclareMathOperator*{\argmax}{arg\,max}

\def\x{{\mathbf x}}
\def\y{{\mathbf y}}
\def\z{{\mathbf z}}
\def\W{{\mathbf W}}

\begin{document}

\pagestyle{headings}
\mainmatter
\def\ECCV18SubNumber{4}  

\title{Bridging machine learning and cryptography in defence against adversarial attacks} 

\titlerunning{Machine learning and cryptography against adversarial attacks}

\authorrunning{O. Taran, S. Rezaeifar \and S. Voloshynovskiy}

\author{Olga Taran \and 	Shideh Rezaeifar \and Slava Voloshynovskiy
\thanks{This work was supported by the SNF project No. 200021-165672.}}
\institute{Computer Science Department,\\ University of Geneva, Geneva, Switzerland
\email{\{olga.taran,shideh.rezaeifar,svolos\}@unige.ch}}

\maketitle

\begin{abstract}
In the last decade, deep learning algorithms have become very popular thanks to the achieved performance in many machine learning and computer vision tasks. However, most of the deep learning architectures are vulnerable to so called \textit{adversarial examples}.
This questions the security of deep neural networks (DNN) for many security- and trust-sensitive domains. The majority of the proposed existing adversarial attacks are based on the differentiability of the DNN cost function. Defence strategies are mostly based on machine learning and signal processing principles that either try to {\em detect-reject} or {\em filter} out the adversarial perturbations and completely neglect the classical cryptographic component in the defence.

In this work, we propose a new defence mechanism based on the second Kerckhoffs's cryptographic principle which states that the defence and classification algorithm are supposed to be known, but not the key. 

To be compliant with the assumption that the attacker does not have access to the secret key, we will primarily focus on a \textit{gray-box} scenario and do not address a \textit{white-box} one. More particularly, we assume that the attacker does not have direct access to the secret block, but (a) he completely knows the system architecture, (b) he has access to the data used for training and testing and (c) he can observe the output of the classifier for each given input.  We show empirically that our system is efficient against most famous state-of-the-art attacks in \textit{black-box} and \textit{gray-box} scenarios.  
\end{abstract}

\section{Introduction}
In the last decade, scientists achieved a big breakthrough in enhancement of functionality  and extension of scope of applications of the DNN. Nowadays, neural networks have become very efficient in many machine-learning tasks. However, despite the remarkable progress the DNN stay vulnerable to adversarial attacks that aim at designing such a perturbation to original samples that, in general, is imperceptible for humans, but it is able to trick the DNN. 
This vulnerability seriously restricts the usage of the DNN in many security- and trust-sensitive domains.

In recent years, researchers proposed a large number of different defence mechanisms. However, the growing number of defences stimulates the invention of even more universal attacks. This is due to the fact that the overwhelming majority of existing attacking algorithms are based on the principle of end-to-end differentiability of the DNN and possibility to add the modification back to the original spatial domain images.

In this paper, we propose a new defence mechanism for the DNN classifiers based on the second Kerckhoffs's cryptographic principle, that states that the fewer secrets the system contains, the higher its safety \cite{massey1993cryptography}. In this regard, the structure of the proposed system is supposed to be known, except the key that is kept secret. This key is used in a security imposing pre-processing block that might be implemented in various ways. 

Based on the best cryptographic practice, one can state that there does not exist any secure algorithm that does not contain any secret unknown to the attacker. From this point, the defences against so called \textit{white-box} attacks will unlikely find practical applications besides some rare exceptions. In this respect, we will primarily focus our attention to a \textit{gray-box} scenario that assumes that the attacker has general knowledge about the structure of the system of interest. However, due to the existence of secret parameters/key, in reasonable time, he is capable neither to discover or to estimate these secret elements, even with the help of modern computational means, nor to train the "bypass" system that would give a sufficiently accurate estimation of the secret parameters/key.

We evaluate our defence mechanism on two standard datasets, namely, MNIST \cite{lecun2009mnist} and Fashion-MNIST \cite{xiao2017fashion} for the \textit{Fast Gradient Sign Method} (FGSM) \cite{goodfellow2014explaining} as the simplest attack case, and attack proposed by N. Carlini and D. Wagner in \cite{carlini2017towards} as the most efficient one for most of the existing adversarial defences mechanisms. 

The main contributions of this paper are:  
\begin{itemize}
\item We analyse the existing state-of-the-art adversarial attacks and most well-known defence algorithms.
\item We present a new defence mechanism for DNN classifiers based on cryptographic principles.
\item We investigate the efficiency of the proposed approach on two standard datasets for several well-known adversarial attacks.
\item We empirically show that a sufficiently simple data-independent transformation, based on a secret key, can serve as a reasonable defence against gradient based adversarial attacks.
\end{itemize}

\subsubsection{Notations.}

We use small bold letters $\x$ to denote a signal that can be represented either in 1D or 2D format, $\x_i$ corresponds to the $i^{\textrm{th}}$ entry of vector $\x$. $E(.)$ and $D(.)$ denotes the \textit{encoder} and \textit{decoder} parts of the DNN classifier, respectively. $P(.)$ indicates the data-independent transformation operator.\\

The remainder of this paper is organised as follows: Section \ref{sec:background} briefly summarizes the basic principle of the DNN classifier and gives the general classification of the existing state-of-the-art attacks against the DNN classifiers as well as a brief classification of the existing defence mechanisms. Section \ref{sec:proposed_approach} introduces the main idea and principles of the proposed defence approach. Section \ref{sec:evaluation} presents the empirical results obtained for the proposed algorithm and their analysis. 
Finally, Section \ref{sec:conclusions} concludes the paper.

\section{Background}
\label{sec:background}

\subsection{Neural Networks}
In general case, it is possible to represent a DNN classifier as a model that consists of two parts: (a) \textit{encoder} or training part and (b) \textit{decoder} or classifier. As an input, the \textit{encoder} takes a multidimensional vector $\x \in \mathbb{R}^{N \times C}$, where $C$ is the number of channels, and outputs a vector $\y \in \mathbb{R}^M$, where in most of the cases, $M$ is equal to the number of classes, and each $\y_j$ is treated as a probability that a given input $\x$ belongs to class $j$. Typically, the \textit{encoder} consists of several nested layers:

\begin{equation}
\y = E(\x) = \sigma_{n} \bigg (\W_{n} \sigma_{n-1} \big ( ...\sigma_1 (\W_1 \x) \big) \bigg), 
\label{eq:eq_1}
\end{equation}

where, at each layer $i$, $\W_i$ corresponds to the model parameters and $\sigma_i$ is an  activation function, usually, non-linear, with $1 \le i \le n$. 

The \textit{decoder} assigns most likely class label based on the result of the \textit{encoder}: 

\begin{equation}
D(\x) = \argmax_{1\le j \le M} E(\x)_j 
\label{eq:eq_2}
\end{equation}
\begin{displaymath}
\;\;\;\;\;\;\;\;\;\; = \argmax_{1\le j \le M} \y_j = \hat{j}.
\end{displaymath}

\subsection{Attacks against DNN classifiers}

Based on the knowledge available to the attacker, the adversarial attacks against the DNN classifiers can be combined into three main groups \cite{yuan2017adversarial}:
\begin{enumerate}
\item \textit{White-box} attacks require the full knowledge of treated model and/or training and test data.
\item \textit{Black-box} attacks, where the attacker (a) has no information about the  structure and parameters of the used model and training data, (b) has possibility to observe the class labels assigned to chosen input like as in case of cryptographic oracle.
\item \textit{Gray-box} attacks, where the attacker (a) knows the system architecture, (b) has access to the data used for training and testing, (c) for each given input can observe the assigned class label, (d) does not have access to or knowledge of the defence mechanism parameters.
\end{enumerate}

In turn, based on the goal of the attacker, the attacks can be \cite{yuan2017adversarial}:
\begin{itemize}
\item \textit{Targeted} that aim at modifying the input in a way that the classifier classifies it as a specified target class. Namely,  for a given target class $t$ such that $D(\x) \ne t$, the goal is to find such a perturbation $\z$ that $D(\x+\z) = t$ \cite{carlini2017towards}.
\item \textit{Non-targeted} that aim at modifying the input in a way that the classifier classifies it incorrectly. Namely, the goal is to find such a perturbation $\z$ that $D(\x+\z) \ne D(\x)$, i.e., the classifier output any wrong class label \cite{carlini2017towards}.
\end{itemize}

It should be pointed out that many adversarial attacks have a transferability property that consists in the fact that adversarial examples trained on one model can be successfully applied to another model with a different architecture of the \textit{encoder} part \cite{yuan2017adversarial}. In general, targeted adversarial examples are much harder to transfer than non-targeted ones.

Due to the limited interest of \textit{white-box} attacks for real-life applications, in our paper we focus preliminary our attention on the \textit{gray-box} targeted and non-targeted attacks that can be also extended to the \textit{black-box} scenario.

\subsubsection{State-of-the-art attacks against DNN classifiers.} 

Without loss of generality, it is possible to group the state-of-the-art adversarial attacks against the DNN classifiers into two main groups:
\begin{enumerate}
\item \textit{Gradient} based attacks. The core principle of which consists in the end-to-end differentiability of many neural network classification systems. \\

This group comprises L-BFGS attack proposed by Szegedy et al. in \cite{szegedy2013intriguing}. This attack is time-consuming due to the used expensive linear search and, as a consequence, is impractical for real-time applications. However, this attack served as a basis for several more successful attacks such as \textit{Fast Gradient Sign Method} (FGSM) \cite{goodfellow2014explaining}. In contrast to L-BFGS, FGSM is fast, but not all the time gives the minimal adversarial perturbation between original and targeted samples. FGSM method has several successful extensions, like FGSM with momentum \cite{dongboosting}, \textit{One-step Target Class Method} (OTCM) \cite{kurakin2016adversarial}, RAND-FGSM algorithm \cite{tramer2017ensemble}, proposed in \cite{kurakin2016adversarial_2} \textit{Basic Iterative Method} (BIM) and \textit{Iterative Least-Likely Class Method} (ILLC), etc. In addition, it should also be mentioned the \textit{Jacobian-based Saliency Map Attack} (JSMA) \cite{papernot2016limitations} and the \textit{DeepFool} approach \cite{moosavi2016deepfool} with its extension \textit{Universal perturbation} \cite{moosavi2017universal}. Moreover, one should note the attack proposed by N. Carlini and D. Wagner in \cite{carlini2017towards}. As it has been shown in many works, like for example in \cite{carlini2017adversarial} and \cite{he2017adversarial}, this attack is among the most efficient ones against many existing defence mechanisms. Finally, A. Athalye et al. in \cite{athalye2018obfuscated} propose Backward Pass Differentiable Approximation technique that aims at avoiding the gradient masking in \textit{white-box} scenario.\\[1cm]

\item \textit{Non-gradient} based attacks\\

The attacks of this group do not require any knowledge of the DNN gradients. The most well-known members of this group are the \textit{Zeroth Order Optimisation} (ZOO) \cite{chen2017zoo} and the \textit{One Pixel Attack} \cite{su2017one}.

\end{enumerate}

\subsection{Defence strategies}
\label{sebsec:defences}
In general, the existing state-of-the-art defence strategies can be classified into four main groups:

\begin{enumerate}
\item \textit{Defence via retraining} \\
The most well-known work in this group is \textit{network distillation} proposed by Papernot et al. in \cite{papernot2016distillation}. Moreover, a significant number of papers were focused on the investigation of the potential of \textit{adversarial retraining}. The main idea behind this retraining is to use adversarial examples in varying degrees during the network training. Representatives of this approach are the works of Goodfellow et al. \cite{goodfellow2014explaining}, Huang et al. \cite{huang2015learning}, Kurakin et al. \cite{kurakin2016adversarial}, Wu et al \cite{wu2017adversarial}, etc. \\

\item \textit{Defence via detection and rejection} \\
Formally, one can distinguish several subclasses, the main ideas behind which can be mixed:
	\begin{enumerate}
		\item Based on integration of \textit{additional DNN}, like for example in the work of Metzen et al. \cite{metzen2017detecting} the original network is augmented by adding an auxiliary "detector" sub-network that aims at distinguishing the original data from data containing adversarial perturbations.
		\item \textit{Statistical} based, for example in \cite{hendrycks2017early} the authors claim that the adversarial perturbations affect in a special way the lower-ranked principal components from PCA. In \cite{li2016adversarial}, the authors perform analysis of the statistics of convolutional layers outputs to detect the adversarial inputs. Feinman et al. in \cite{feinman2017detecting} analyse Bayesian uncertainty available in dropout neural networks, etc.
		\item Based on analysis of \textit{DNN properties}, such as \cite{katz2017reluplex} and \cite{katz2017towards} where the authors make their decision by extending the simplex algorithm to support the non-convex ReLUs activation function and consider the neural network as a whole, without any simplifying assumptions.\\
	\end{enumerate}
	
\item \textit{Defence via input pre-processing} \\
In general case, defence is achievable through filtering and removal of modifications introduced to original images, like for example the denoising used in \cite{meng2017magnet} or recently proposed in \cite{lee2018defensive}.\\

\item \textit{Defence via regeneration} \\
The key point behind this approach consists in the assumption that the adversarial examples can be mapped back to the manifold of the original clean data via regeneration. For example, Gu et al. in \cite{gu2014towards} propose a \textit{deep contractive autoencoder} that is a variant of the classical autoencoder with an additional penalty increasing the robustness to adversarial examples. Meng et al. in \cite{meng2017magnet} introduce \textit{MagNet}, which combines the detector and regeneration networks.

\end{enumerate}

Despite the big variety of the existing defence mechanisms, there are still several important open issues: 
 \begin{itemize}
\item In the vast majority of cases, the attacker knows exactly the same amount of information or can easily learn any complementary information about the defence strategy. Therefore, the attacker can relatively easy bypass the defence mechanism.
\item Usually, there is no information advantage for the defender over the attacker.
\item In general, no cryptographic principles are used by the defender.
\end{itemize}

\section{Proposed approach}
\label{sec:proposed_approach}

In this paper, we propose for the first time up to our best knowledge, a defence strategy using a cryptographic formulation. This strategy is based on the next principles:
 \begin{itemize}
\item {\em Information advantage of the defender over attacker}: we consider a party consisting of: encoder $E(.)$ and classifier (decoder) $D(.)$ that share a common secret, i.e., secret key that is used in one of the blocks. This party is a defender that plays against the attacker, who does not know the secret key.
\item {\em Data-independence of security imposing transformation}. The transformation used in the security imposing block  should be based on a secret key and should be data-independent to avoid possibility to be learned from the training dataset. The entropy of the secret key should be at least as high as the entropy of the signal. In addition, it is desirable, but not absolutely necessary, that the security imposing transformation is non-differentiable. Non-differentiability makes the proposed approach end-to-end non-differentiable and provides the additional difficulties for the attacker.
\item {\em Protection of "internal variables" via the assumed protocol}. To be compliant with most of practical requirements about the deployment of AI we will assume that the attacker has access to the output of the classifier, but can not observe the internal variables of the network to avoid the access to the secret key or its easy learning protocols. This assumption is reasonable assuming that the recognition is done on protected servers or special devices or chips, which is typical in domains as for example biometric or digital watermarking applications facing similar concerns. 
\end{itemize}

One can imagine the extension of similar principles to other applications such as digital forensics, device identification, etc., facing similar deficiencies.\\

\begin{figure}[t!]
	\centering
	\includegraphics[width=0.75\linewidth]{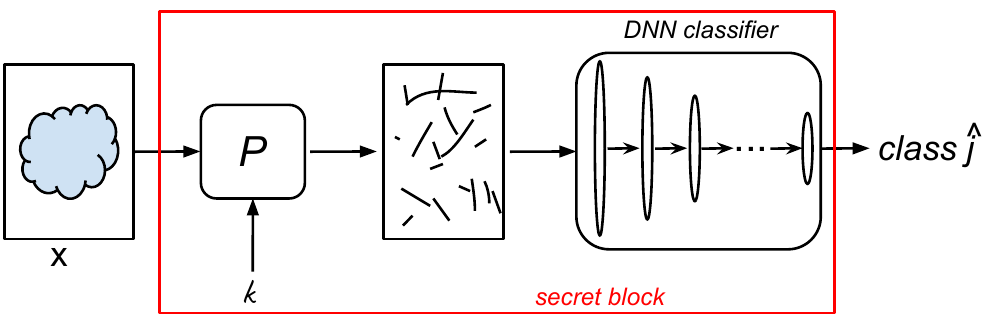} 
	\caption{Principal scheme of the proposed algorithm: input and output are observable variables for the attacker, the secret block architecture is known, but the inner parameters and variables are not observable.}
	\label{fig:schema}
\end{figure}

The generalized scheme of the proposed approach is illustrated in Fig. \ref{fig:schema} and is quite simple. At first, an input signal $\x \in \mathbb{R}^{N \times C}$ is fed through a specially designed transform block $P$, where the secret key $k \in \mathcal{K}$ is used. After that, the result comes to the input of the DNN classifier. In general, the architecture of block $P$ and the DNN classifier is supposed to be public. Following the classification given in Section \ref{sebsec:defences}, the proposed defence can be formally associated to the class of \textit{defences via input pre-processing}. However, in the proposed approach, the pre-processing block $P$ does not assume any filtering or artefact removing. The main requirement to this block consists in the fact that it should be data-independent transformation $P(.)$ based on a secret key that in addition, preferably, but not necessary, can be non-invertible and non-differentiable. Mathematically, for the given input signal $\x \in \mathbb{R}^{N \times C}$ the classification can be defined as:

\begin{equation}
D(\x) = \argmax_{1 \le j \le M} E(P(\x, k))_j,
\label{eq:eq_3}
\end{equation}
where $k$ is a known secret key. 

\subsubsection{Requirements to secret elements.}
Following the second Kerckhoffs' cryptographic principle \cite{massey1993cryptography} we assume that all details of our algorithms are publicly known and available to the attacker besides the key. At the same time, taking into account the capabilities of modern computing systems, keeping a key secret can be not so trivial. We assume that if the attacker would have access to the output of transform block $P$, he can easily discover the key $k$ or make the defence system end-to-end differentiable in his attack by simply replacing the block $P$ by a trained differentiable mapper or applying the Backward Pass Differentiable Approximation technique recently proposed by A. Athalye et al. in \cite{athalye2018obfuscated}. In this regard, we assume the use of some standard measures that restricts the access to the internal results and parameters of the proposed system.

Moreover, in order to protect the system against brute force attacks, theoretically the entropy of the key is supposed to be at least as high as those of the input signal. In practice, it is supposed that the key is used correctly meaning that it is random and is unique (never reused) for each application instance. In our case, we will assume that each classifier has its own key as for example each self-driving car equipped by a DNN based recognition system would have an individual classifier parametrized by $k$. Obviously, in this case, the process of training is more complex in comparison with one common non-secret classifier.  

\section{Evaluation}
\label{sec:evaluation}

\subsection{Datasets}

\begin{figure}[t!]
	\centering
	\includegraphics[width=1\linewidth]{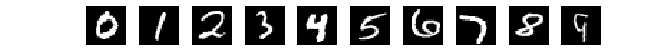} 
	\includegraphics[width=1\linewidth]{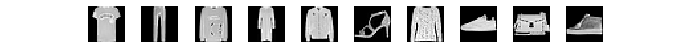}
	\caption{Examples of original images from each class from MNIST (top line) and Fashion-MNIST (bottom line) datasets.}
	\label{fig:datasets_examples}
\end{figure}

Our evaluation strategy starts with more simple examples and evolves to more complex ones. As a simple dataset we used the MNIST set of hand-written digits \cite{lecun2009mnist} that contains 10 classes, 60 000 training and 10 000 test grayscale images of the size $28 \times 28$. Fashion-MNIST set \cite{xiao2017fashion} serves as a dataset with more complex and diverse structure of objects. Similarly to MNIST, Fashion-MNIST consists of 10 classes, 60 000 training and 10 000 test grayscale images of the size $28 \times 28$. Examples of images from each dataset are illustrated in Fig. \ref{fig:datasets_examples}.

In case of both datasets, we used 55 000 examples for training and 5 000 for validation from the train set, and the first 1 000 samples from the test set for testing on the original and attacked data.

\subsection{Base-line attacks}

Guided by the same principle as when selecting databases, we selected two well-known algorithms of adversarial attacks, namely, the \textit{Fast Gradient Sign Method} (FGSM) \cite{goodfellow2014explaining} as a simplest case, and the attack proposed by N. Carlini and D. Wagner in \cite{carlini2017towards} (that will be referred to as CW) as the most efficient for most of existing adversarial defences mechanisms.

\subsubsection{Fast Gradient Sign Method.}
The FGSM method determines for each pixel of input signal $\x$ the direction in which the intensity of pixel should be changed by using the gradient of the loss function \cite{goodfellow2014explaining}:
\begin{equation}
\x^t = \x + \epsilon \cdot sign(\nabla_{\x}J(\W, \x, t)), 
\end{equation}
where $t$ is a target class, $\x^t$ is the adversarial input, $J(.)$ is the loss function of the DNN classifier, $\W$ denotes the parameters of the model and $\epsilon$ is a small constant that controls the level of distortions. 

In general, the FGSM attack was developed to be fast, but it does not always produce the minimal required distortions.

\subsubsection{CW attack.} 
In contrast to the FGSM, the CW attack proposed in \cite{carlini2017towards} aims at finding for a given input $\x \in \mathbb{R}^{N \times C}$ the adversarial example $\x^t = \x + \boldsymbol\delta$ with a minimum possible distortion $\boldsymbol\delta$:

\begin{equation}
\begin{array}{cll}
\min_{\delta} & \| \boldsymbol\delta \|_p + c \cdot f(\x + \boldsymbol\delta) &\\
s.t           & 0 \le \x_i + \delta_i \le 1             & \forall i=1, ..., (N \times C),\\
\end{array}
\end{equation}
where $c > 0$ is a suitable chosen constant, $\boldsymbol\delta$ is the desired distortion, $f(.)$ is a new objective function, such that $D(\x + \boldsymbol\delta) = t$, if and only if $f(\x + \boldsymbol\delta) \le 0$, $\|.\|_p$ is a $\ell_p$ norm defined as:

\begin{displaymath}
\|\boldsymbol\delta\|_p = \Bigg ( \sum_{i=1}^{N \times C} |\delta_i|^p  \Bigg)^{\frac{1}{p}}.
\end{displaymath} 

The authors in \cite{carlini2017towards} investigate several objective functions $f$ and as the most effective they propose:

\begin{equation}
f(\x^t) = \max \big( \max\{Z(\x^t)_j: j\ne t \} - Z(\x^t)_t, -\kappa \big),
\end{equation}  
where $Z$ is the result of the network before the last activation function that, in case of classification, usually is a \textit{softmax}, according to the equation (\ref{eq:eq_1}), $E(\x) = \sigma_n(Z(\x))$ and $\kappa$ is a constant that controls the confidence. 

\subsection{Empirical results and analysis}

\begin{table}[t!]

\centering
\begin{minipage}[t]{0.45\linewidth}
\renewcommand*{\arraystretch}{1.2}
\caption{Architecture of the DNN classifiers used for the FGSM attack.}
\label{tab:fgsm_dnn_architecture}
\begin{tabular}{lK{2cm}K{2cm}K{2cm}} \hline
Layer   & \# filters & kernel size \\ \hline
Conv2D $\;$  & 64 & $8 \times 8$ \\ 
ReLu    & & \\ 
Conv2D  & 128 & $6 \times 6$ \\ 
ReLu    & & \\ 
Conv2D  & 128 & $5 \times 5$ \\
ReLu    & & \\ 
Flatten & & \\ 
Dense   & & 10 \\ \hline
\end{tabular}

\end{minipage}
$\;\;\;\;$
\begin{minipage}[t]{0.45\linewidth}
\renewcommand*{\arraystretch}{1.2}
\caption{Architecture of the DNN classifiers used for the CW attack.}
\label{tab:cw_dnn_architecture}
\begin{tabular}{lK{2cm}K{2cm}K{2cm}} \hline
Layer & \# filters & kernel size \\ \hline
Conv2D  & 32/64 & $3 \times 3$ \\ 
ReLu    & & \\ 
Conv2D  & 32/64 & $3 \times 3$ \\ 
ReLu    & & \\ 
MaxPool & & $2 \times 2$ \\ 
Conv2D  & 64/128 & $3 \times 3$ \\ 
ReLu    & & \\ 
Conv2D  & 64/128 & $3 \times 3$ \\ 
ReLu    & & \\ 
MaxPool & & $2 \times 2$ \\ 
Flatten & & \\ 
Dense   & & 200/256 \\
ReLu    & & \\ 
Dropout & & \\ 
Dense   & & 200/256 \\
ReLu    & & \\ 
Dense   & & 10 \\ \hline
\end{tabular}

\end{minipage}
\end{table}


As it has been described in Section \ref{sec:proposed_approach}, the proposed approach consists of two parts: (1) a security imposing block $P$ and (2) a standard DNN classifier. 

In order to follow the best practices of reproducible research, in our experiments we used the FGSM attack implementation from the CleverHans python library\footnote{\url{https://github.com/tensorflow/cleverhans}}. The structure of the DNN classifier used for the generation of this kind of the adversarial examples is given in Table \ref{tab:fgsm_dnn_architecture}. For the generation of the CW adversarial examples we used the code provided by the authors of this attack\footnote{ \url{https://github.com/carlini/nn\_robust\_attacks}}. The structure of the corresponding DNN classifier is shown in Table \ref{tab:cw_dnn_architecture}. In this way we investigate the applicability of proposed defence approach to the different architectures of DNN classifiers.

As it has been indicated in Section \ref{sec:proposed_approach}, the fundamental requirement to the block $P$ is its data-independence and the presence of the secret key $k$. In general case, one can choose $P(.)$ from a broad family of transformations. In our experiments, in order to validate our theory, we considered the simplest case of $P(.)$ that is a standard random permutation based on the secret key $k$ . For the simplicity of our analysis, we have assumed that the length of the key is equal to the dimensionality of the input data. It should be noted that in the general case it does not ensure that $H(k) \ge H(\x)$, where $H(.)$ denotes entropy. However, since all images are normalized between 0 and 1 and posses correlation while the key is initialized from i.i.d Gaussian noise $\mathcal{N}(0, 1)$, our assumption is reasonably satisfied. This allows us to reduce to a minimum the number of possible parameters that could affect the results and comprehension of whether the proposed approach can serve as a natural defence against the adversarial attacks.

Before starting the analysis of the obtained results, it should be pointed out that our goal was not to obtain the state-of-the-art classification accuracy on the chosen datasets, but we aim at investigating the possibilities of the proposed defence idea from the point of view of existence of simple yet efficient mechanisms based on proven by practice cryptographic principles for the defence of the DNN classifiers against the adversarial attacks.

The obtained experimental results for the discussed attacks on the chosen datasets are given in Table \ref{tab:classification_error}, where "classical classifier" corresponds to the standard DNN classifiers without any defence and with the parameters indicated in Tables \ref{tab:fgsm_dnn_architecture} - \ref{tab:cw_dnn_architecture}. The term "classifier on permuted data" corresponds to the proposed approach with the same standard DNN classifiers. 

\begin{table}[t!]
\renewcommand*{\arraystretch}{1.25}
\centering
\caption{Classification error (\%) on the first 1 000 test samples}
\label{tab:classification_error}
\begin{tabular}{|K{2.5cm}|K{2cm}|K{2cm}|K{2cm}|K{2cm}|} \hline
\multirow{ 2}{*}{Attack} & \multicolumn{2}{|c|}{Classical classifier} & \multicolumn{2}{|c|}{Classifier on permuted data} \\ \cline{2-5}
& original & attacked & original & attacked \\ \hline

\multicolumn{5}{|c|}{\textit{MNIST}}\\ \hline
CW $\ell_2$      & 1.00 & 100.00 & 3.00 & 8.64 \\ \hline
CW $\ell_0$      & 1.00 & 100.00 & 3.00 & 14.53 \\ \hline
CW $\ell_\infty$ & 1.00 & 99.99  & 3.00 & 12.24 \\ \hline
FGSM             & 1.00 & 92.10  & 1.40 & 18.00 \\ \hline

\multicolumn{5}{|c|}{\textit{Fashion MNIST}}\\ \hline
CW $\ell_2$      & 7.50 & 100.00 & 11.50 & 12.12 \\ \hline
CW $\ell_0$      & 7.50 & 100.00 & 11.50 & 13.48\\ \hline
CW $\ell_\infty$ & 7.50 & 99.90  & 11.50 & 12.55 \\ \hline
FGSM             & 8.60 & 60.60	& 11.20 & 27.50  \\ \hline
\end{tabular}
\end{table}

According to Table \ref{tab:classification_error}, in the case of the simple MNIST dataset and original non-attacked input samples, the proposed defence leads to an insignificant decrease of classification accuracy for both tested architectures of the DNN classifiers with respect to the one without defence. As for adversarial attacks, then it should be noted that such a simple defence strategy decreases the classification error from 92 - 100\% to 9 - 18\% for all types of considered attacks.  In case of the Fashion-MNIST dataset, the observed tendency is the same. Thus, the obtained results demonstrate a great potential of the proposed concept and show that even a simple random permutation in the defence block $P$ can serve as a quite promising defence. 

From the point of view of security, the attacker can try to guess the secret key via brute force. However, one should take into account that in the proposed approach the length of the used key is at least as the length of the input signal. In case of the (Fashion-) MNIST dataset the size of the input signal is 784, which is sufficiently large, to make the brute force attack practically infeasible. Moreover, taking into account the fact that attacker does not have access to the internal results of the system, he will not be able to train a "bypass" mapper between the input of the system and output of the transform block $P$ in order to avoid the necessity to know the secret key or to apply the Backward Pass Differentiable Approximation technique.

\section{Conclusions}
\label{sec:conclusions}

In this work our main focus was the defence of the DNN classifiers in \textit{gray-box} scenario as the most suitable for the real-life applications based on the best cryptographic practices. We briefly discussed the state-of-the-art adversarial attacks and most well-known defence algorithms and proposed our vision on a possible classification for each direction. We presented a new defence mechanism based on cryptographic principles that can be applied to many existing DNN classifiers. It should be noted that our goal was not to reach a state-of-the-art classification accuracy on the considered datasets. Our empirical findings showed a quite promising potential of the proposed idea from the point of view of existence of simple yet efficient mechanisms based on proven by practice cryptographic principles for the defence of the DNN classifier against the gradient based adversarial attacks. 

As a future work we aim at investigating other data-independent transformations that can be used in the proposed algorithm and examining the behaviour of our defence on the more complex datasets, like for example CIFAR-10 \cite{krizhevsky2014cifar} and  on the non-gradient based adversarial attacks. 

%
%
\bibliographystyle{splncs}
\bibliography{0004}
\end{document}